%% file: lp_lc.tex
\documentclass{JAC2003}
\usepackage{graphicx}

\input{definitions}

\begin{document}

\input{title.tex}

%% \title{Physics at Future Linear Colliders}

%% \author{K. M\"onig}

%% \address{
%% DESY, Zeuthen, Germany and LAL, Orsay France
%% \\E-mail: klaus.moenig@desy.de}

%% \twocolumn[\maketitle\abstract{
%% This article summarises the physics at future linear colliders. It will be
%% shown that in all studied physics scenarios a 1\,TeV linear collider in
%% addition to the LHC will enhance our knowledge significantly and helps to 
%% reconstruct the model of new physics nature has chosen.
%% }]

\section{Introduction}%1
Most physicists agree that the International Linear Collider, ILC, should be
the next large scale project in high energy physics\cite{consensus}. The ILC
is an $\ee$ linear collider with a centre of mass energy of $\sqrt{s} \le 500
\GeV$ in the first phase, upgradable to about $1\TeV$\cite{jonathan}.  The
luminosity will be ${\cal L} \approx 2-5 \cdot 10^{34} \lunit$ corresponding
to $200 - 500 \fbi$/year. The electron beam will be polarisable with a
polarisation of ${\cal P} = 80-90\%$.

In addition to this baseline mode there are a couple of options whose
realisation depends on the physics needs. With relatively little effort also
the positron beam can be polarised with a polarisation of $40 - 60 \%$. The
machine can be run on the Z resonance producing $>10^9$ hadronically decaying
Z bosons in less than a year or at the W-pair production threshold to measure
the W-mass to a precision around $6\MeV$ (GigaZ). 
The ILC can also be operated as an ${\rm e}^- {\rm e}^-$
collider. With much more effort one or both beams can be brought into
collision with a high power laser a few mm in front of the interaction point
realising a $\gamma\gamma$ or ${\rm e} \gamma$ collider with a photon energy
of up to 80\% of the beam energy. 

At a later stage one may need an $\ee$
collider with $\rts = 3-5 \TeV$. Such a collider may be realised in a two-beam
acceleration scheme (CLIC). Extensive R\&D for such a machine is currently
going on\cite{frank}.

ILC will run after LHC\cite{gigi,fabiola} has taken already several years of
data. However the two machines are to a large extend complementary. The LHC
reaches a centre of mass energy of $\rts = 14 \TeV$ leading to a very high
discovery range. However not the full $\rts$ is available due to parton
distributions inside the proton
($\rts_{\rm eff} \sim 3 \TeV$).
The initial state is unknown and the proton remnants disappear in the beampipe
so that energy-momentum conservation cannot be employed in the analyses.
There is a huge QCD background and thus not all processes are visible.

ILC has with its $\sqrt{s} \le 1\TeV$ a lower reach for direct discoveries.
However the full $\rts$ is available for the primary interaction and the
initial state is well defined, including its helicity.  The full final state
is visible in the detector so that energy-momentum conservation also allows
reconstruction of invisible particles.
Since the background is small, basically all processes are visible at the ILC.

The LHC is mainly the ``discovery machine'' that can find new particles up to
the highest available energy and should show the direction nature has taken. On
the contrary ILC is the ``precision machine'' that can reconstruct the
underlying laws of nature. Only a combination of the LHC reach with the ILC
precision is thus able to solve our present questions in particle physics.

Better measurement precision can not only improve existing knowledge but
allows to reconstruct completely new effects. For example Cobe discovered the
inhomogeneities of the cosmic microwave background but only the precision of
WMAP allowed to conclude that the universe is flat. As another example, from
the electroweak precision measurements before LEP and SLD one could verify that
the lepton couplings to the Z were consistent with the Standard Model
prediction but only the high precision of LEP and SLD could predict the Higgs
mass within this model.

The ILC has a chance to answer several of the most important questions in
particle physics. Roughly ordered in the chances of the ILC to find some
answers they are:
\begin{itemize}
\item How is the electroweak symmetry broken?
  The ILC can either perform a precision study of the Higgs system or see
  first signs of strong electroweak symmetry breaking.
\item What is the matter from which our universe is made off?  ILC has a high
  chance to see supersymmetric dark matter, also some other solutions like
  Kaluza Klein dark matter might give visible signals.
\item Is there a common origin of forces?
  Inside supersymmetric theories the unification of couplings as well as of
  the SUSY breaking parameters can be checked with high precision.
\item Why is there a surplus of matter in the universe?
  Some SUSY models of baryogenesis make testable predictions for the ILC. Also
  CP violation in the Higgs system should be visible.
\item How can gravity be quantised?
  The ILC is sensitive to extra dimensions up scales of a few TeV
  and tests of unification in SUSY may give a hint
  towards quantum gravity at the GUT scale.
\end{itemize}

\section{The Top Quark Mass and why we need it}

\begin{figure}[htb]
  \centering
  \includegraphics[width=\linewidth]{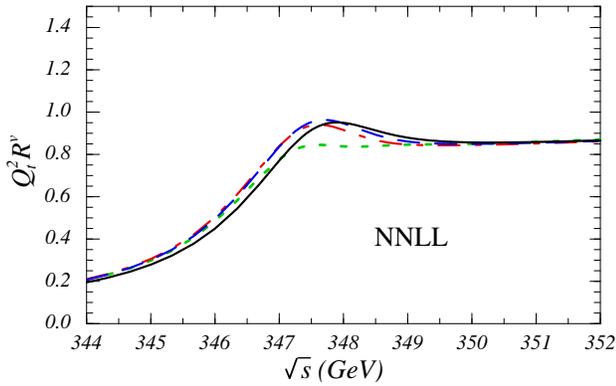}
  \caption[]{Top pair cross section using the NNLL pole mass for different
  values of the top velocity parameter\cite{Hoang:2001mm}. }
  \label{fig:sect}
\end{figure}

ILC can measure the top mass precisely from a scan of the $\ttb$ threshold.
With the appropriate mass definition the cross section near threshold is well
under control\cite{Hoang:2001mm} (see fig.~\ref{fig:sect}). With a ten-point
scan an experimental precision of $\Delta \MT  = 34 \MeV $ and 
$\Delta \Gamma_{\rm t} =  42 \MeV $ is possible\cite{Martinez:2002st}, so
that, including theoretical uncertainties 
$\Delta \MT(\overline{{\rm MS}}) \approx 100 \MeV $ can be reached.

\begin{figure*}[htb]
  \centering
  \includegraphics[width=0.48\linewidth]{MWSW04.cl.eps}
  \includegraphics[width=0.46\linewidth]{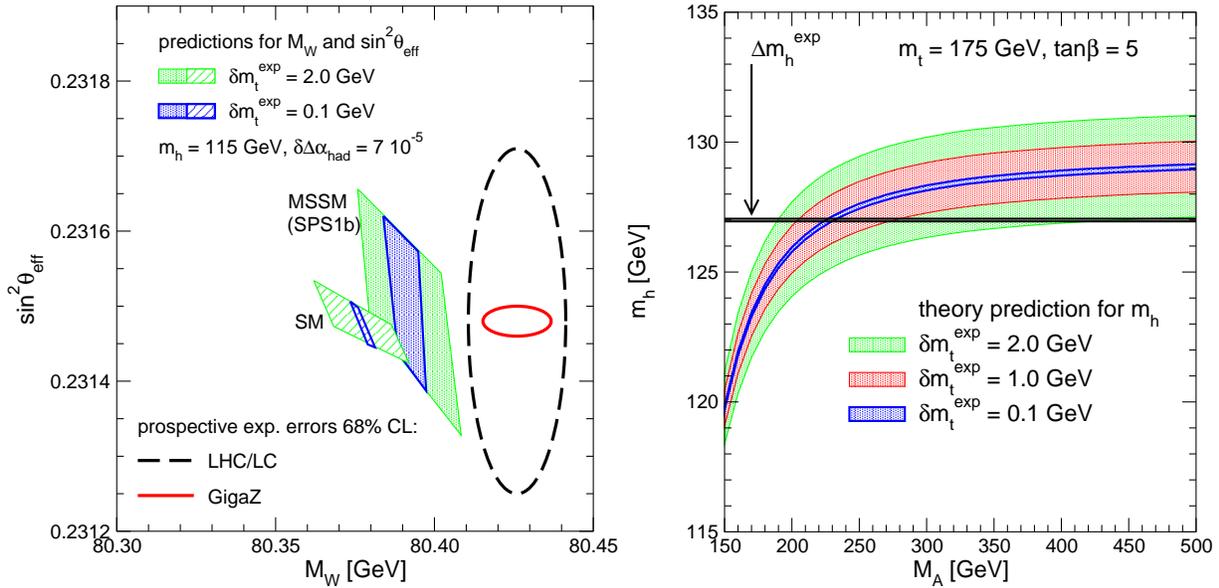}
  \caption{Required top mass precision for the interpretation of the
  electroweak precision data (left) and for the MSSM Higgs system (right).}
  \label{fig:mtprec}
\end{figure*}

A precise top mass measurement is needed in many applications. The
interpretation of the electroweak precision data after GigaZ needs a top mass
precision better than 2\,GeV (fig.~\ref{fig:mtprec} left) and the
interpretation of the MSSM Higgs system even needs a top mass precision of
about the same size as the uncertainty on the Higgs mass 
(fig.~\ref{fig:mtprec} right)\cite{georg_top}.
Also the interpretation of the WMAP cosmic microwave data in terms so the MSSM
needs a good top mass precision in some regions of the parameter 
space\cite{wmap_top}.

\section{Higgs Physics and Electroweak Symmetry Breaking}

If a roughly Standard Model like Higgs exists, it will be found by the LHC.
However the ILC has still a lot to do to figure out the exact model and to
measure its parameters. If only one Higgs exists it can be the Standard Model,
a little Higgs model or the Higgs can be mixed with a Radion from extra
dimensions. If two Higgs doublets exist it can be a general two Higgs doublet
model or the MSSM. However the Higgs structure may be even more complicated
like in the NMSSM with an additional Higgs singlet or the top quark can play a
special role as in little Higgs or top-colour models. In all cases there maybe
only one Higgs visible at LHC that looks Standard-Model like, but the
precision at ILC can distinguish between the models.

\begin{figure}[htb]
  \centering
   \includegraphics[width=0.9\linewidth,bb=40 5 531 507,clip]{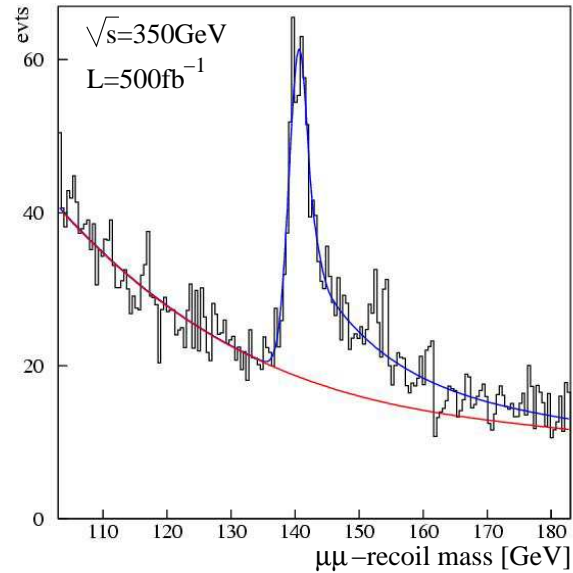}
  \caption{Measurement of $\ee \rightarrow HZ$ from the $\mumu$ recoil mass.}
  \label{fig:hrec}
\end{figure}

The Higgs can be identified independent from its decay mode using the $\mumu$
recoil mass in the process $\ee \rightarrow HZ$ with $Z\rightarrow \mumu$
(see fig.~\ref{fig:hrec})\cite{jeanclaude}. The cross section of this process
is a direct measurement of the HZZ coupling and it gives a bias free
normalisation for the Higgs branching ratio measurements.
Together with the coss section of the WW fusion channel 
($\ee \rightarrow \nu \nu H$) this allows for a model independent
determination of the Higgs width and its couplings to W, Z,
b-quarks, $\tau$-leptons, c-quarks and gluons on the $1-5\%$ 
level\cite{phys_tdr}.

\begin{figure}[htb]
  \centering
  \includegraphics[width=0.9\linewidth]{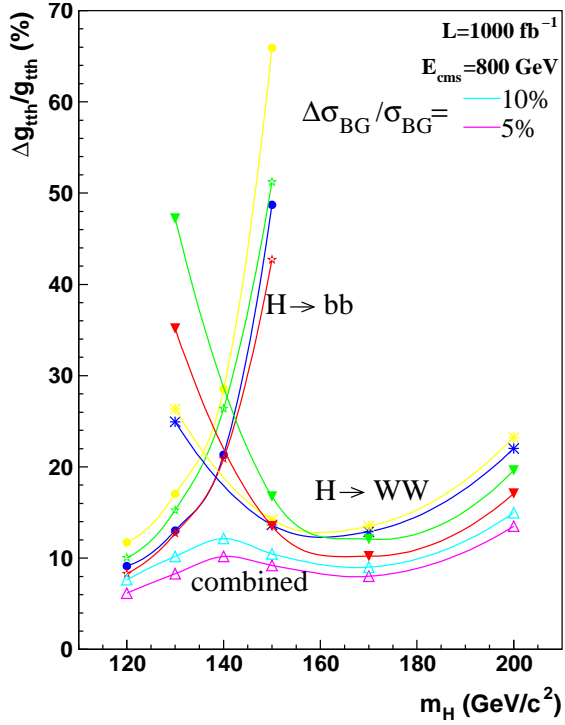}
  \caption{Expected precision of the $\ttb H$ Yukawa coupling as a function of
  the Higgs mass.}
  \label{fig:htt}
\end{figure}

At higher energies the $\ttb H$ Yukawa coupling can be measured from the
process $\ee \rightarrow \ttb H$ where the Higgs is radiated off a t-quark. At
low Higgs masses, using $H \rightarrow \bb$, a precision around 5\% can be
reached. For higher Higgs masses, using $H \rightarrow WW$, 10\% accuracy will
be possible (see fig.~\ref{fig:htt})\cite{gay}.

\begin{figure}[htb]
  \centering
  \includegraphics[width=\linewidth]{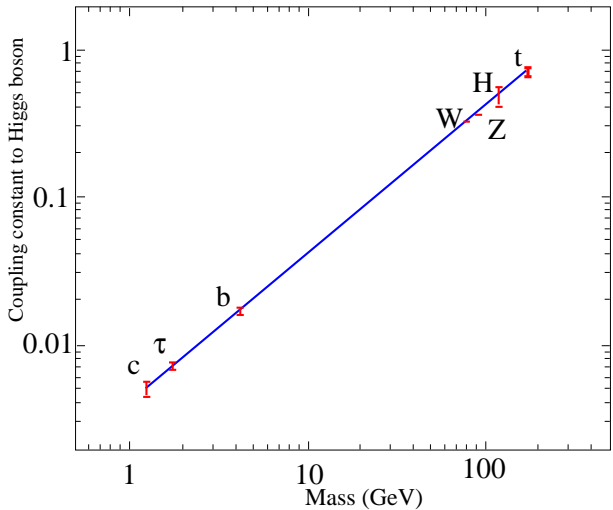}
  \caption{Higgs-particle coupling and expected uncertainty as a function
    of the particle mass.}
  \label{fig:hmcoup}
\end{figure}

If the Higgs is not too heavy the triple Higgs self-coupling can be measured
to around 10\% using the double-Higgs production channels $\ee \rightarrow
{\rm ZHH}$ and $\ee \rightarrow \nu \bar{\nu}{\rm HH}$\cite{satoru}. As shown
in fig.~\ref{fig:hmcoup} all these Higgs coupling measurements allow to show,
that the Higgs really couples to the mass of the particles\cite{satoru}.

\begin{figure}[htb]
  \centering
  \includegraphics[width=\linewidth]{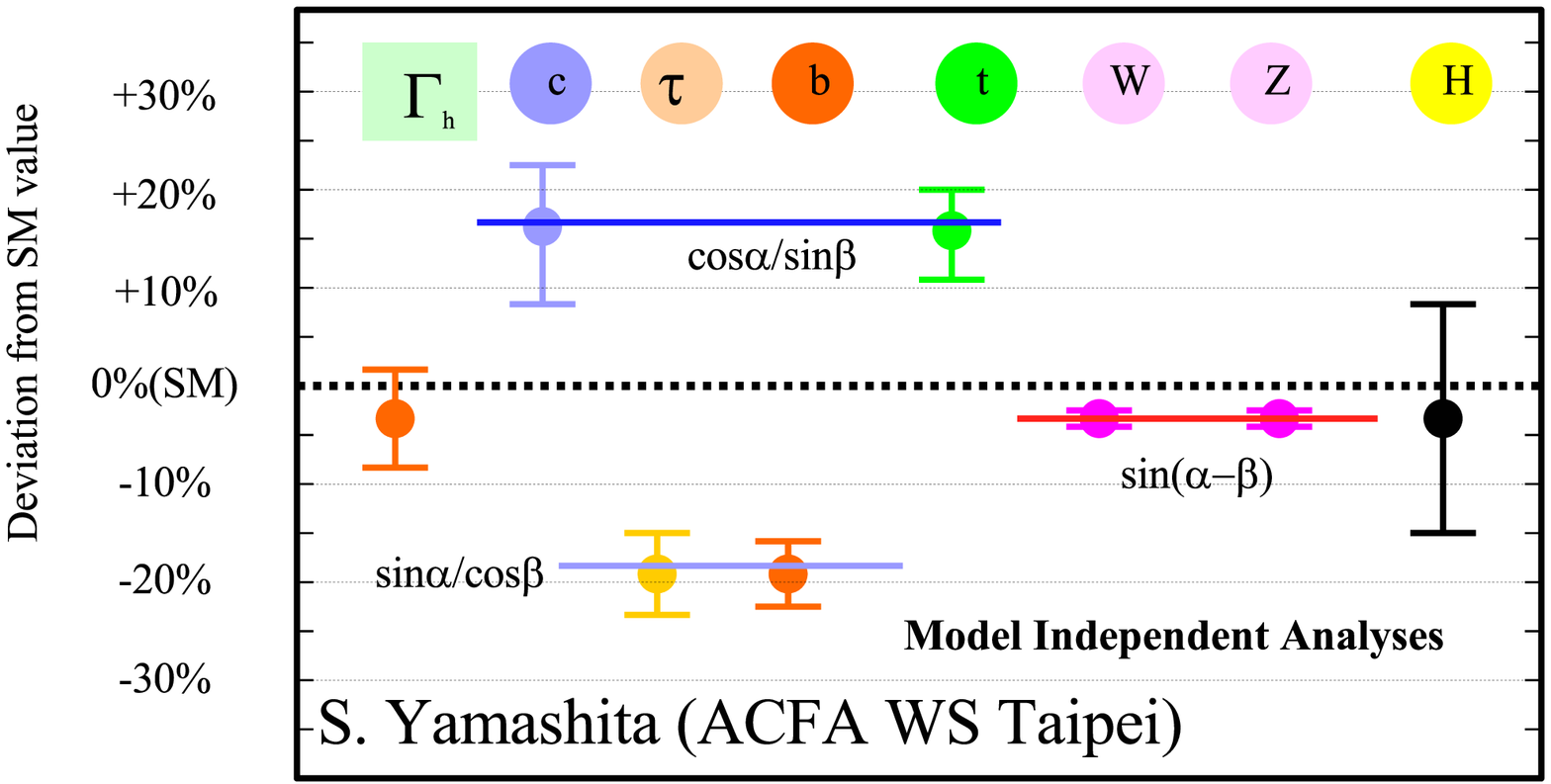}
  \includegraphics[width=\linewidth]{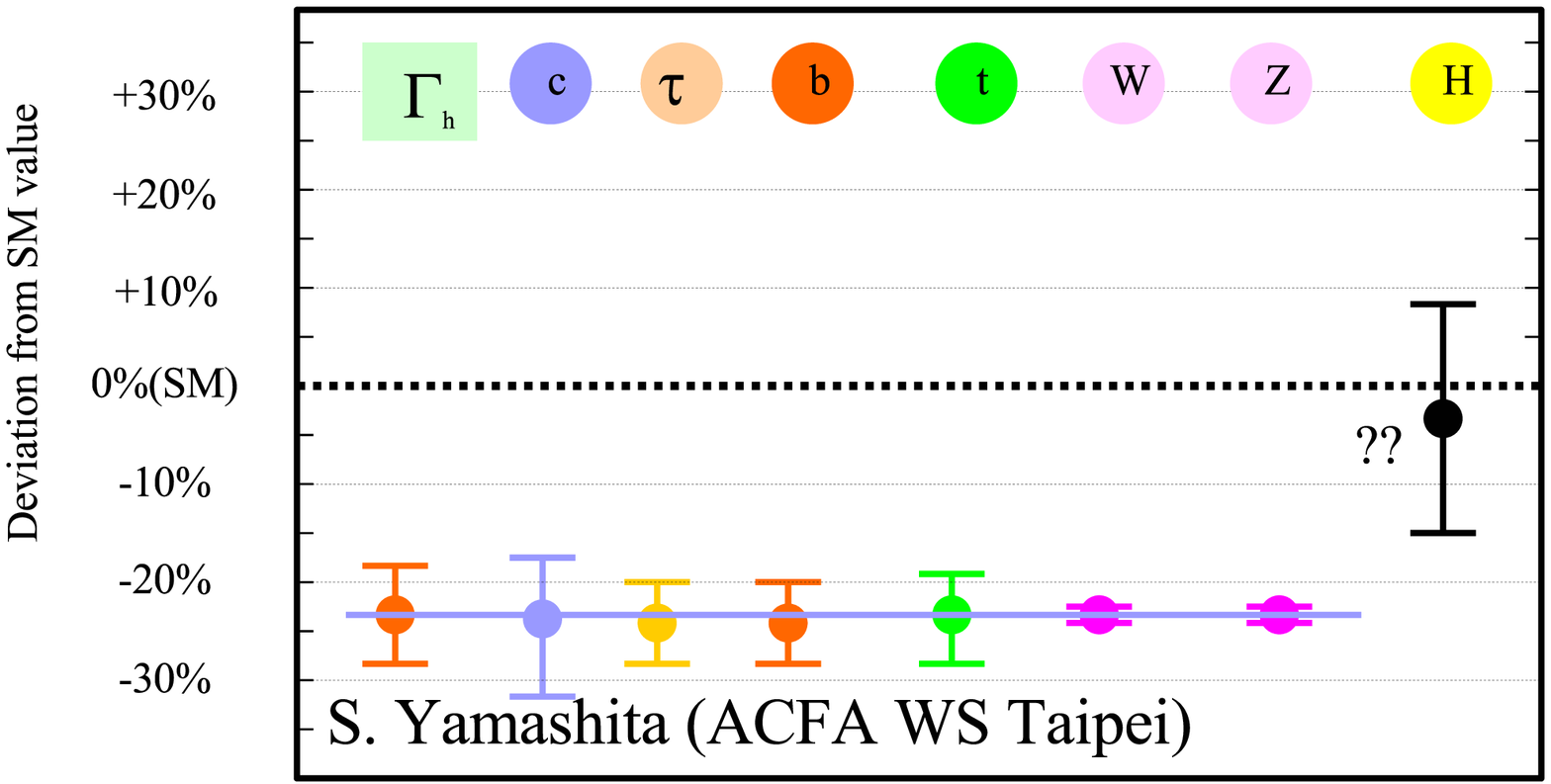}
  \includegraphics[width=\linewidth]{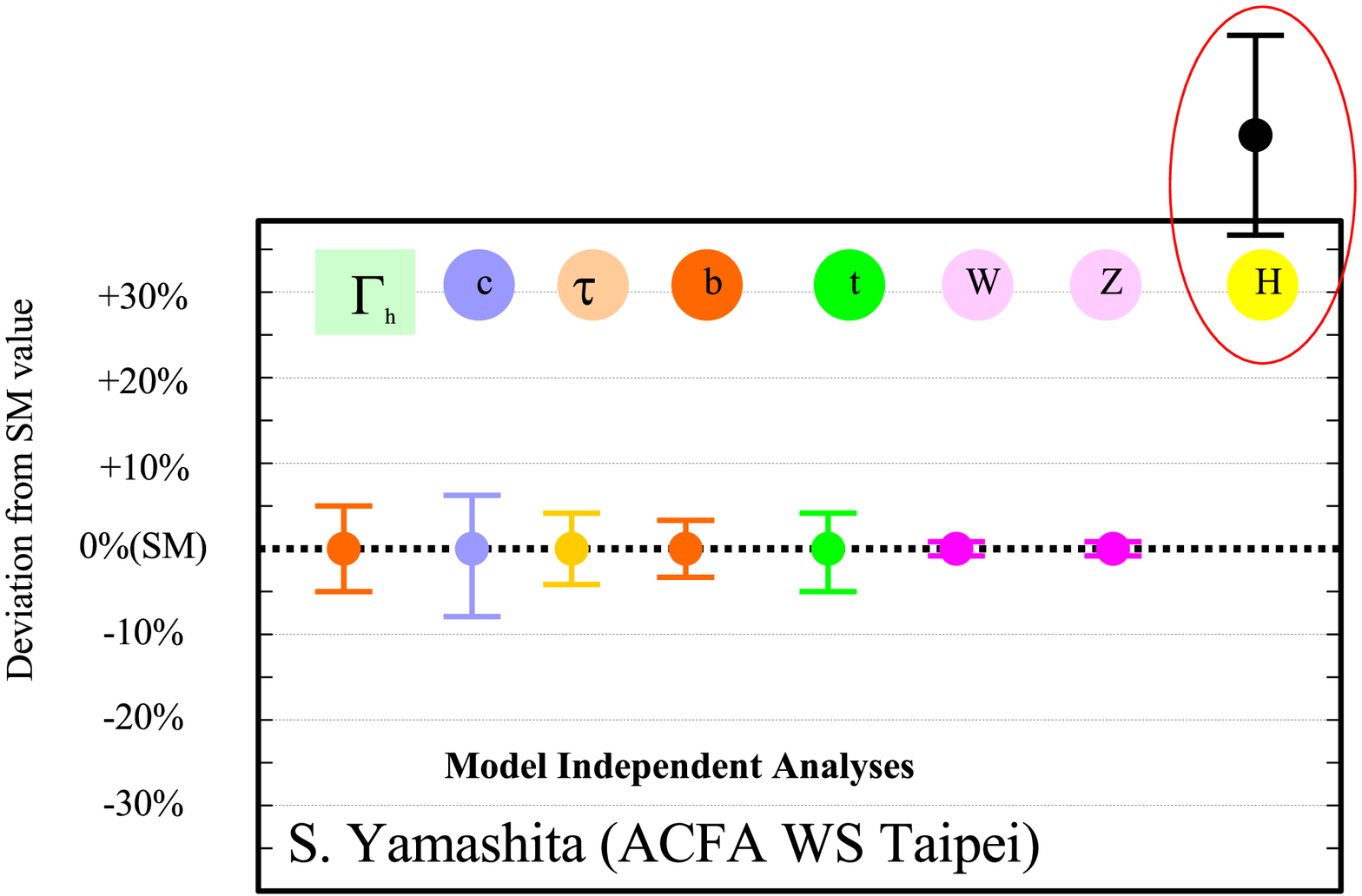}
  \caption[]{Deviation of the Higgs couplings from the Standard Model together
    with the expected ILC precision for a two Higgs doublet model (upper), a
    model with Higgs-Radion mixing (middle) and a model incorporating
    baryogenesis\cite{bghhh} (lower).}
  \label{fig:sath}
\end{figure}

These measurements present a powerful tool to test the model
from which the Standard Model Higgs arises. Figure \ref{fig:sath}\cite{satoru}
shows possible deviations of the Higgs couplings from the the Standard Model
prediction together with the expected uncertainties for a two Higgs doublet
model, a model with Higgs-Radion mixing and a model incorporating
baryogenesis\cite{bghhh}. In all cases the ILC allows to separate clearly
between the Standard Model and the considered one.

\begin{figure}[htb]
  \centering
  \includegraphics[width=0.82\linewidth]{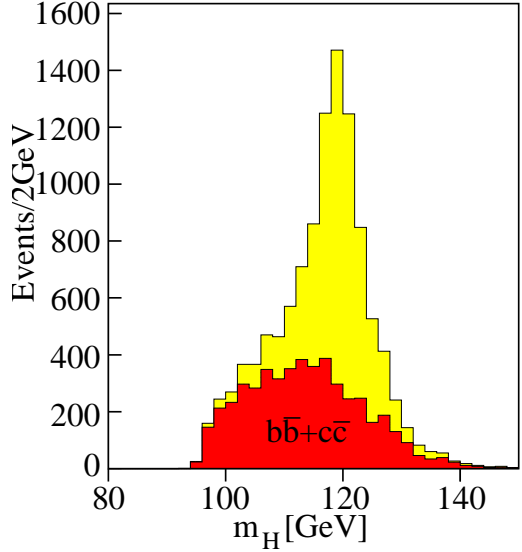}
  \caption[]{$\bb$ mass spectrum in the $\gamma \gamma \rightarrow H$ analysis
    after all cuts\cite{hphot1}.}
  \label{fig:ggh}
\end{figure}

\begin{figure}[htb]
  \centering
  \includegraphics[height=\linewidth,angle=-90]{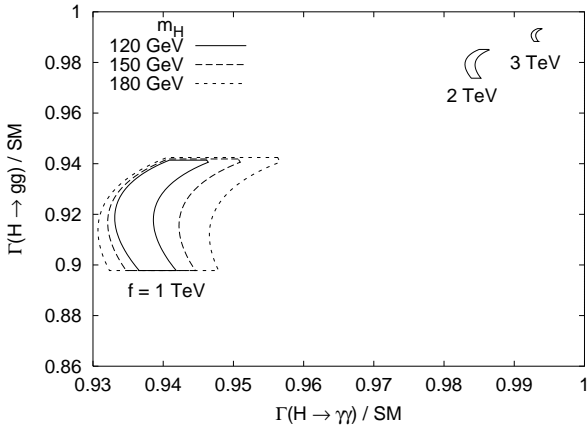}
  \caption{Possible deviations of Higgs loop decays from the Standard Model
    prediction in little Higgs models.}
  \label{fig:hloop}
\end{figure}

Further information can be obtained from loop decays of the Higgs, namely $H
\rightarrow gg$ and $H \rightarrow \gamma \gamma$. Loop decays probe the Higgs
coupling to all particles, also to those that are too heavy to be produced
directly. The Higgs-decay into gluons probes the coupling to all coloured
particles which is completely dominated by the top-quark in the Standard
Model. The one to photons is sensitive to all charged particles, dominantly
the top quark and the W-boson in the SM. The partial width $\Gamma ( H
\rightarrow gg )$ can be measured on the 5\% level from Higgs decays in $\ee$.
The photonic coupling of the Higgs can be obtained from the Higgs production
cross section at a photon collider (see
fig.~\ref{fig:ggh})\cite{hphot1,hphot2}. The loop decays of the Higgs are
sensitive to the model-parameters in many models. As an example figure
\ref{fig:hloop} shows the expected range of couplings within a little Higgs
model\cite{hlittle}.

\subsection{Heavy SUSY-Higgses}
\begin{figure}[htb]
  \centering
  \includegraphics[width=0.93\linewidth,bb=14 8 295 280,clip]{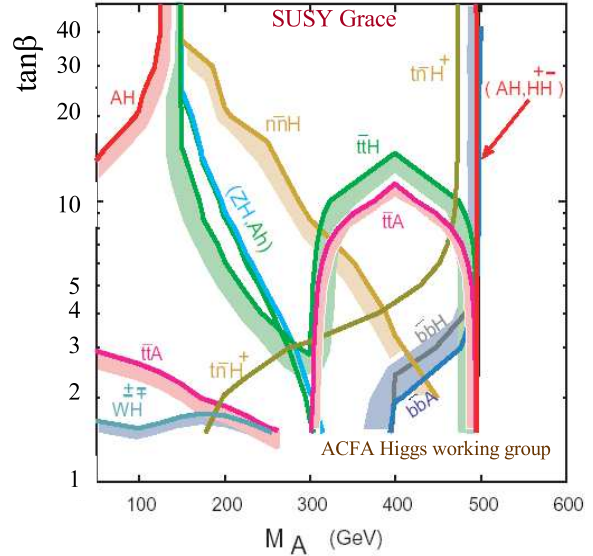}
  \caption{Visibility of heavy SUSY Higgses at ILC ($\rts = 1 \TeV$).}
  \label{fig:HAreach}
\end{figure}

In the relevant parameter range of the MSSM the heavy scalar, H, the
pseudoscalar, A, and the charged Higgses ${\rm H}^\pm$ are almost degenerate in
mass and the coupling ZZH vanishes or gets at least very small. At the ILC they
are thus pair-produced, either as HA or ${\rm H}^+ {\rm H}^-$ and the cross
section depends only very little on the model parameters. All states a
therefore visible basically up to the kinematic limit $m(H) < \rts/2$.
As shown in figure \ref{fig:HAreach}\cite{satoru} at least one of the heavy
states should be visible in another channel in most of the parameter
space. The additional channels serve as redundancy and can be used to
measure model parameters.

\begin{figure}[htb]
  \centering
  \includegraphics[width=0.95\linewidth,bb=19 23 562 463, clip]{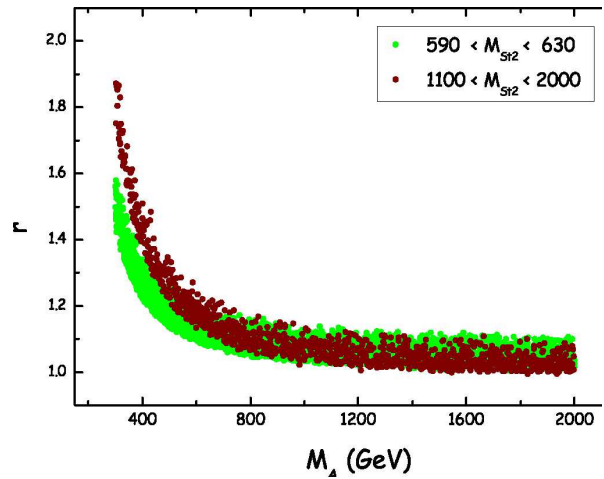}
  \caption{$BR(h \rightarrow \bb)/BR(h \rightarrow WW)$ MSSM/SM within the
  SPS1a scenario as a function of $\MA$.}
  \label{fig:MAind}
\end{figure}

In addition to the direct searches the precision branching ratio measurements
of the light Higgs can give indications of the H and A mass. Figure
\ref{fig:MAind} shows the ratio of branching ratios 
$BR(h \rightarrow \bb)/BR(h \rightarrow WW)$ of the MSSM relative to the
Standard Model as a function of $\MH$\cite{Desch:2004cu}. The width of the
band gives the uncertainty from the measurement of the MSSM parameters. Up to A
masses of a few hundred GeV one can get a good indication of $\MA$.

\begin{figure}[htb]
  \centering
  \includegraphics[width=0.95\linewidth,bb= 1 13 519 516]{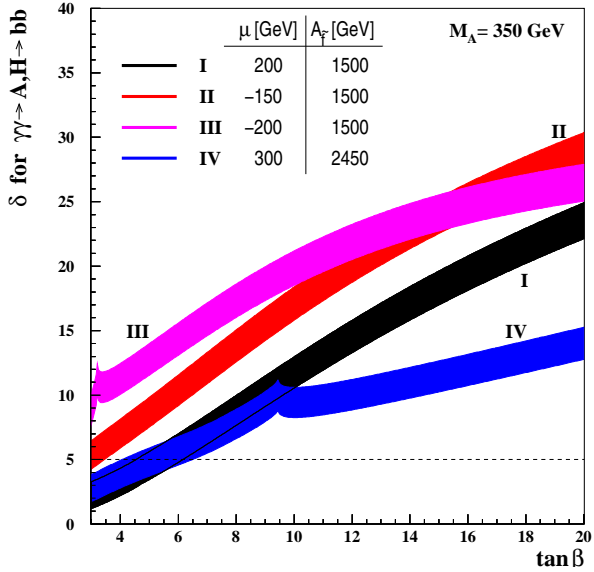}
  \caption{Sensitivity of the $\gamma \gamma$ collider to heavy MSSM Higgses.
    ($\MA = 350 \GeV$, $\rts_{ee} = 500 \GeV$, $M_2 = 200 \GeV$,
    $M_{\tilde{f}} = 1000 \GeV$)}
  \label{fig:ggHA}
\end{figure}

Another possibility to find the heavy SUSY Higgses is the photon collider.
Since Higgses are produced in the s-channel the maximum reach is twice the
beam energy corresponding to $0.8 \rts_{ee}$. Figure \ref{fig:ggHA} shows the
expected sensitivity in one year of running for $\MA = 350 \GeV$, $\rts_{ee} =
500 \GeV$ and different SUSY parameters\cite{ggHA}.  In general H and A are
clearly visible, however due to the loop coupling of the $\gamma$ to the Higgs
the sensitivity becomes model dependent.

\section{Supersymmetry and Dark Matter}

Supersymmetry (SUSY) is the best motivated extension of the Standard Model.
Up to now all data are consistent with SUSY, however also with the pure
Standard Model.  Contrary to the SM, SUSY allows the unification of couplings
at the GUT scale and, if R-parity is conserved, SUSY offers a perfect dark
matter candidate.  If some superpartners are visible at the ILC they will be
discovered by the LHC in most part of the parameter space.  However many tasks
are left for the ILC in this case.  First the ILC has to confirm that the
discovered new states are really superpartners of the Standard Model
particles.  Then it has to measure as many of the $>100$ free parameters as
possible in a model independent way which allows to check if grand unification
works and to get an idea by which mechanism Supersymmetry is broken.  If
Supersymmetric particles are a source of dark matter the ILC has to measure
their properties.

\begin{figure}[htb]
  \centering
  \includegraphics[width=\linewidth,bb=19 37 592 593,clip]{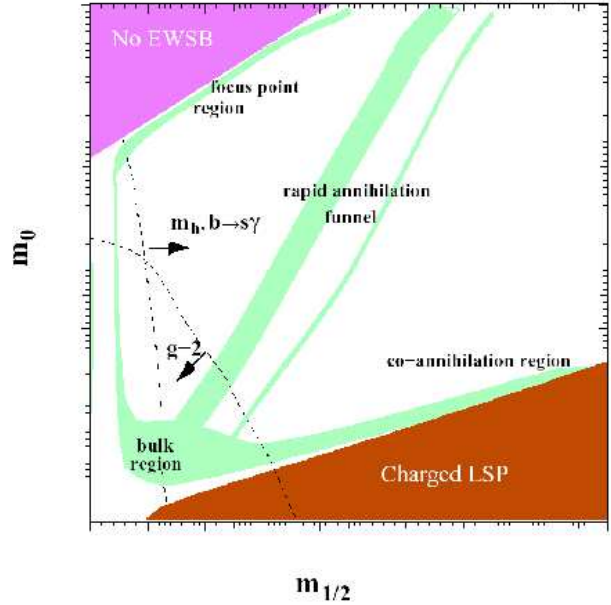}
  \caption{Dark matter allowed regions of mSUGRA.}
  \label{fig:dmmsugra}
\end{figure}

Within the minimal supergravity model (mSUGRA) the parameter space can be
strongly restricted requiring that the abundance of the lightest neutralino,
which is stable in this model, is consistent with the dark matter density
measured by WMAP.  Figure \ref{fig:dmmsugra} shows the allowed region in a
pictorial way\cite{alpgdm}. In the so called ``bulk region'' all superpartners
are light and many are visible at the LHC and the ILC. In the ``coannihilation
region'' the mass difference between the lightest neutralino,
$\tilde{\chi}^0_1$, and the lighter stau, $\tilde{\tau}_1$, is very small so
that the $\tilde{\tau}_1$-decay particles that are visible by the detector have
only a very small momentum. In the ``focus point region'' the
$\tilde{\chi}^0_1$ gets a significant Higgsino component enhancing its
annihilation cross section.  This leads to relatively heavy scalars, probably
invisible at the ILC and the LHC. Other regions, like the ``rapid annihilation
funnel'' are characterised by special resonance conditions, like $2
m(\tilde{\chi}^0_1) \approx \MA$, increasing the annihilation rate. All these
special regions tend to be challenging for both machines.

\begin{figure}[htb]
  \centering
  \includegraphics[width=\linewidth,bb=40 53 552 405,clip]{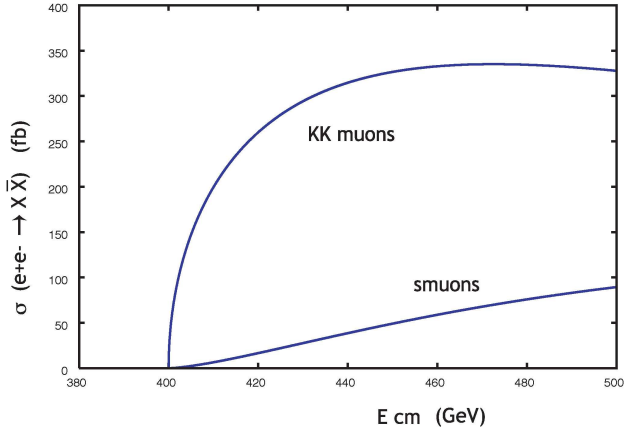}
  \caption{Threshold behaviour of smuon production and the production of
    Kaluza Klein excitations of the muon.}
  \label{fig:kksthresh}
\end{figure}

\begin{figure}[htb]
  \centering
  \includegraphics[width=0.9\linewidth,bb=25 2 300 270]{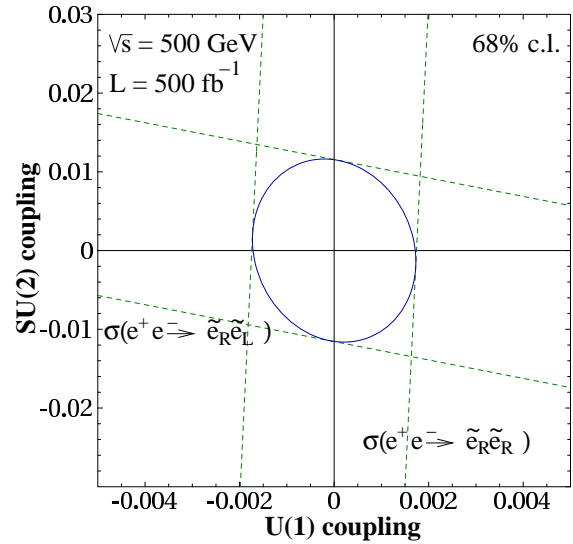}

  \caption{Measurement of the SU(2) and U(1) coupling of the selectron at the 
    ILC.}
  \label{fig:scoup}
\end{figure}

After new states consistent with SUSY have been discovered at the LHC, the ILC
can check, if it is really Supersymmetry. As an example
fig.~\ref{fig:kksthresh} shows the threshold behaviour of smuon production and
the production of Kaluza Klein excitations of the muon\cite{pesvic}. There is
no problem for the ILC to distinguish the two possibilities. Figure
\ref{fig:scoup} shows the expected precision of the measurement of the SU(2)
and U(1) coupling of the selectron\cite{ayrescoup}. The agreement with the
couplings of the electron can be tested to the percent to per mille level.

\subsection{SUSY in the bulk region}

\begin{figure}[htb]
  \centering
  \includegraphics[width=\linewidth,bb=0 67 525 559,clip]{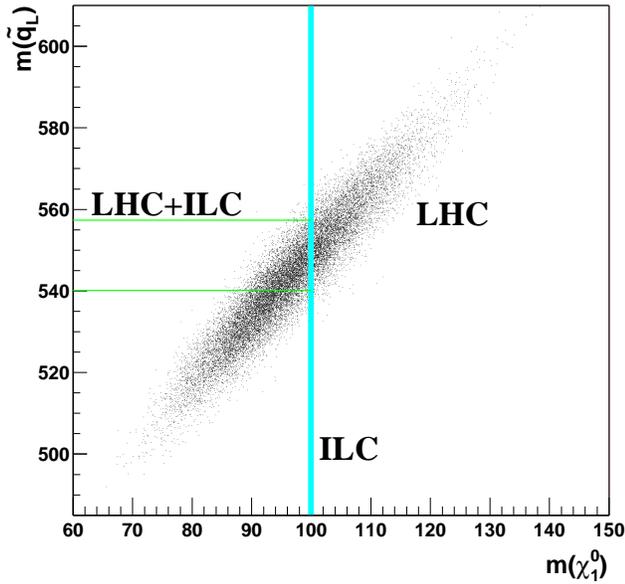}
  \caption{Correlation between $m(\tilde{\chi}^0_1)$ and $m(\tilde{q})$
    measurements at LHC together with $m(\tilde{\chi}^0_1)$ from ILC.}
  \label{fig:lhcmqmc} 
\end{figure}

An often studied benchmark point in the bulk region is the SPS1a
scenario\cite{sps}. In this scenario all sleptons, neutralinos and charginos
are visible at ILC and and in addition squarks and gluinos at the LHC.  The
LHC can measure mass differences pretty accurately, but has difficulties to
measure absolute masses. The ILC, however can measure absolute masses with
good precision, including the one of the $\tilde{\chi}^0_1$. Table
\ref{tab:susy} shows the expected precision for the mass measurements for the
LHC and ILC alone and for the combination\cite{lclhc}.  In many cases the
combination is significantly better than the LHC or even the ILC alone.
As an example figure \ref{fig:lhcmqmc} shows the correlation between the
squark mass and the $\tilde{\chi}^0_1$ mass from LHC together with
$m(\tilde{\chi}^0_1)$ from ILC\cite{lclhc}.  The improvement in
$m(\tilde{q})$ is evident.

\begin{table*}[htb]
\begin{tabular}{|l|cccc||l|cccc|}
\hline
 & $m_{\rm SPS1a}$ & LHC & ILC & LHC+ILC &
 & $m_{\rm SPS1a}$ & LHC & ILC & LHC+ILC\\
\hline
\hline
$h$  & 111.6 & 0.25 & 0.05 & 0.05 &
$H$  & 399.6 &      & 1.5  & 1.5  \\
$A$  & 399.1 &      & 1.5  & 1.5  &
$H+$ & 407.1 &      & 1.5  & 1.5  \\
\hline
$\chi_1^0$ & 97.03 & 4.8 & 0.05  & 0.05 &
$\chi_2^0$ & 182.9 & 4.7 & 1.2   & 0.08 \\ 
$\chi_3^0$ & 349.2 &     & 4.0   & 4.0  &
$\chi_4^0$ & 370.3 & 5.1 & 4.0   & 2.3 \\
$\chi^\pm_1$ & 182.3  & & 0.55 & 0.55 &
$\chi^\pm_2$ & 370.6  & & 3.0  & 3.0 \\
\hline
$\tilde{g}$ &  615.7 & 8.0 &  & 6.5 & & & & & \\
\hline
$\tilde{t}_1$ & 411.8 &     &  2.0  & 2.0 & & & & & \\
$\tilde{b}_1$ & 520.8 & 7.5 &       & 5.7 &
$\tilde{b}_2$ & 550.4 & 7.9 &       & 6.2 \\
\hline
$\tilde{u}_1,\tilde{c}_1$ &  551.0 & 19.0 & & 16.0 &
$\tilde{u}_2,\tilde{c}_2$ &  570.8 & 17.4 & &  9.8 \\
$\tilde{d}_1,\tilde{s}_1$ &  549.9 & 19.0 & & 16.0 &
$\tilde{d}_2,\tilde{s}_2$ &  576.4 & 17.4 & &  9.8 \\
%$\tilde{s}_1$ &  549.9 & 19.0 & & 16.0 &
%$\tilde{s}_2$ &  576.4 & 17.4 & &  9.8 \\
%$\tilde{c}_1$ &  551.0 & 19.0 & & 16.0 &
%$\tilde{c}_2$ &  570.8 & 17.4 & &  9.8 \\
\hline
$\tilde{e}_1$    & 144.9    & 4.8 & .05 & 0.05 &
$\tilde{e}_2$    & 204.2    & 5.0 & 0.2  & 0.2  \\
$\tilde{\mu}_1$  & 144.9    & 4.8 & 0.2  & 0.2  &
$\tilde{\mu}_2$  & 204.2    & 5.0 & 0.5  & 0.5  \\
$\tilde{\tau}_1$ & 135.5    & 6.5 & 0.3  & 0.3  &
$\tilde{\tau}_2$ & 207.9    &     & 1.1  & 1.1  \\
$\tilde{\nu}_e$  & 188.2    &     & 1.2  & 1.2  & & & & & \\
\hline
\end{tabular}
  \centering
  \caption{Expected precision of mass measurements at LHC and ILC in the SPS1a
  scenario.}
  \label{tab:susy}
\end{table*}

\begin{figure}[htb]
  \centering
  \includegraphics[width=\linewidth,bb=20 3 551 379]{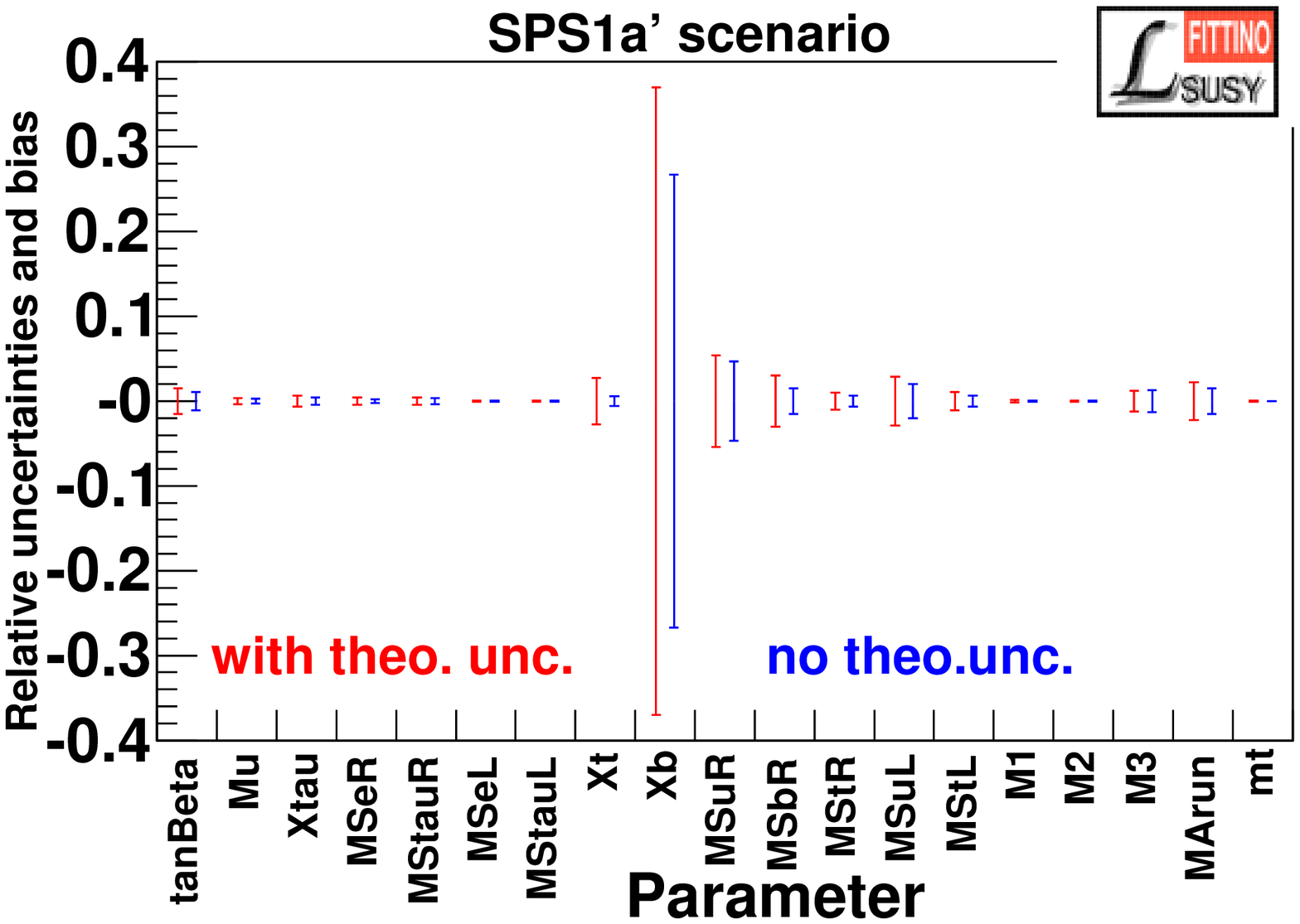}
  \caption{Low energy SUSY breaking parameters from a fit to the LHC and
    ILC results.}
  \label{fig:fittino}
\end{figure}

With these inputs it is then possible to fit many of the low energy SUSY
breaking parameters in a model independent way. Figure \ref{fig:fittino} shows
the result of this fit to the combined ILC and LHC results for the SPS1a
scenario\cite{fittino}. Most parameters can be measured on the percent level.

\begin{figure}[p]
  \centering
  \includegraphics[width=\linewidth]{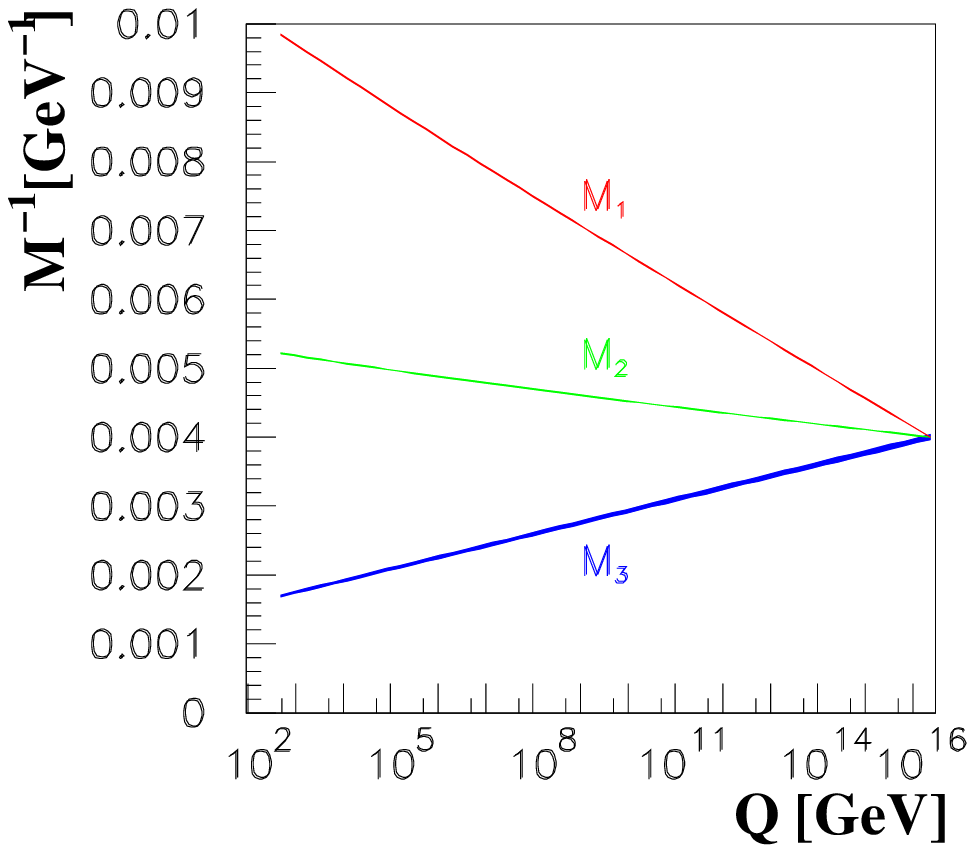}
  \includegraphics[width=\linewidth]{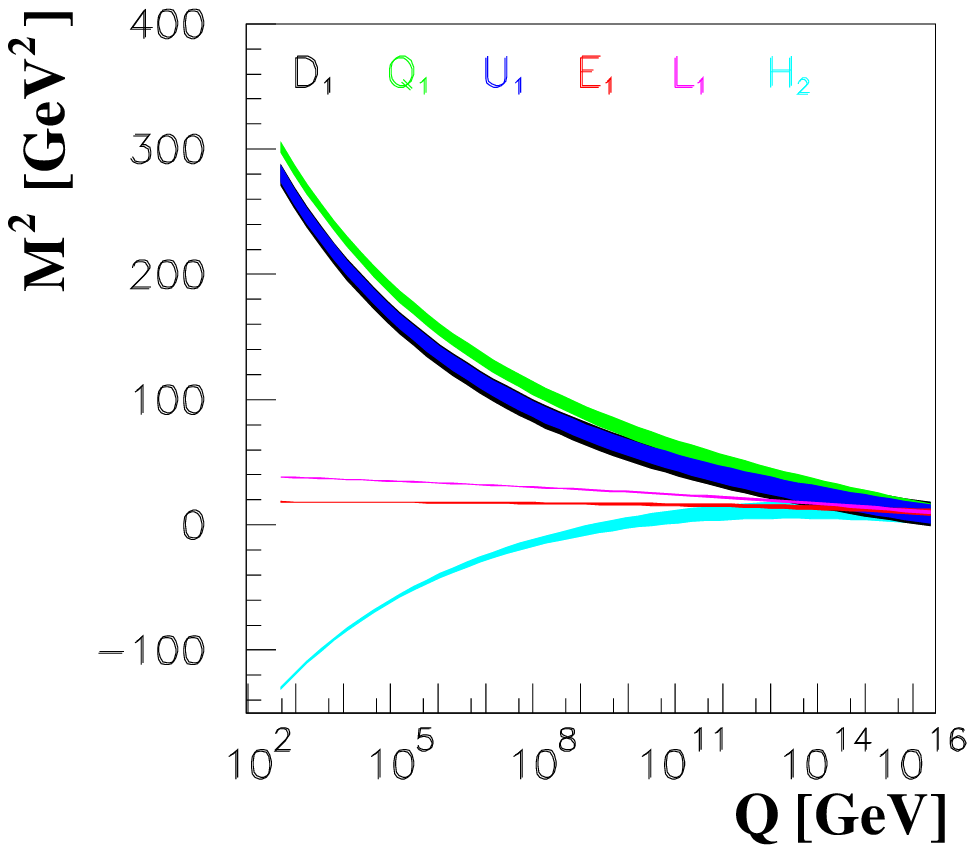}
  \includegraphics[width=0.8\linewidth]{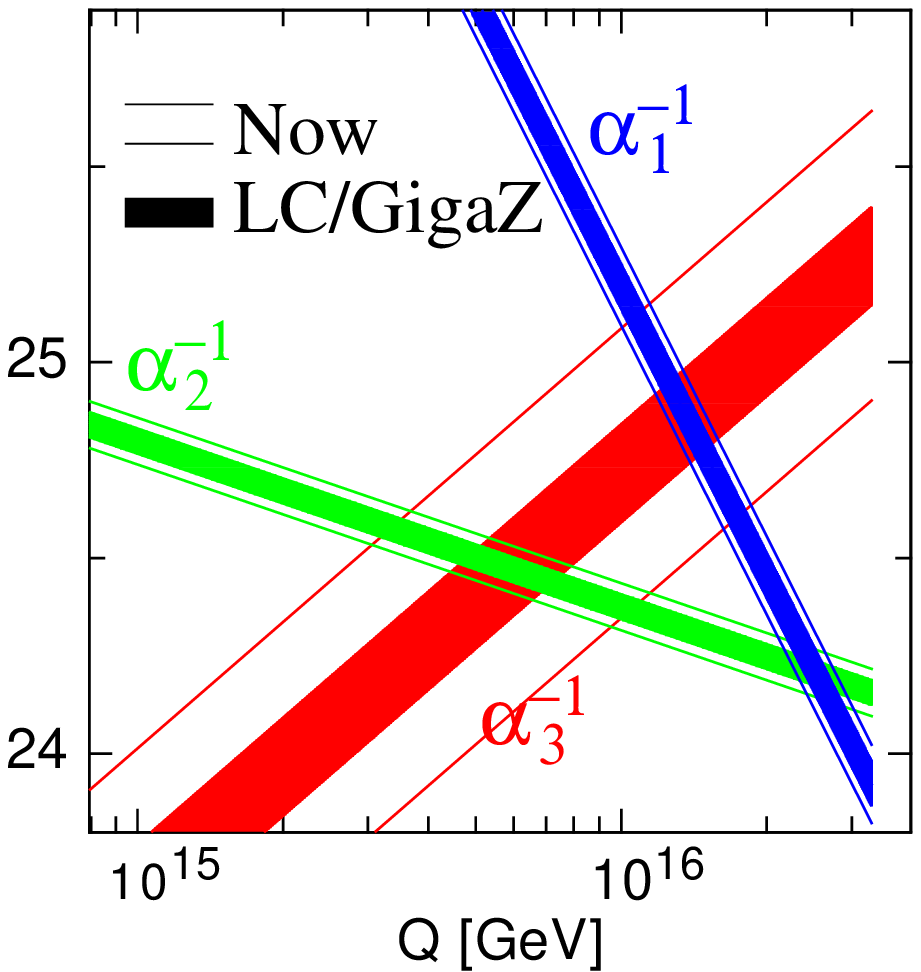}
  \caption{Extrapolation of the gaugino and slepton mass parameters and of 
    the coupling constants to the GUT scale.}
  \label{fig:susygut}
\end{figure}

These parameters can then be extrapolated to high scales using the
renormalisation group equations to check grand unification\cite{susygut}.
Figure \ref{fig:susygut} shows the expected precision for the gaugino and
slepton mass parameters and for the coupling constants.

\subsection{Reconstruction of dark matter}

\begin{figure}[htb]
  \centering
  \includegraphics[width=\linewidth]{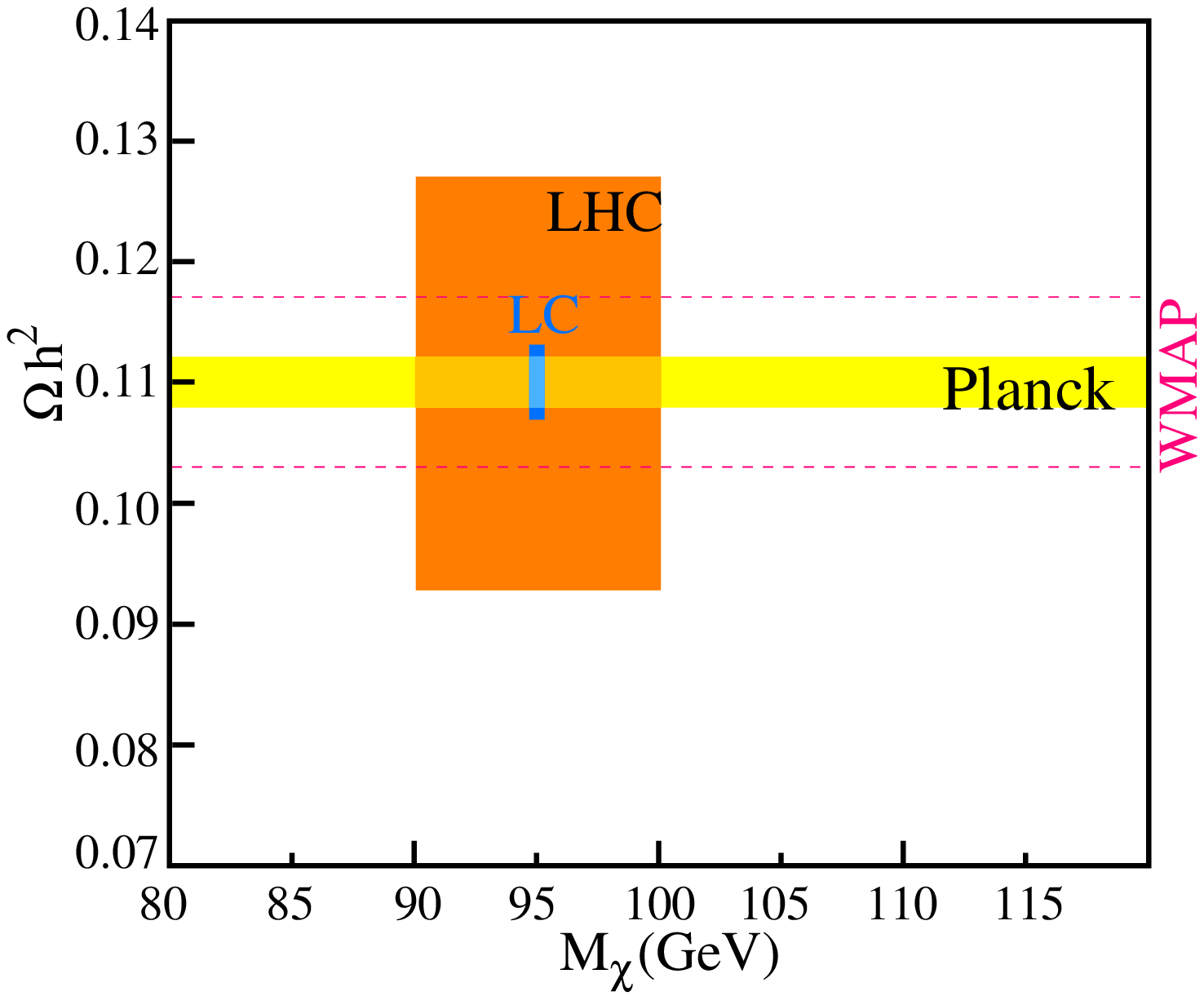}
  \caption{Projected precision of the dark matter density in the 
    coannihilation region from WMAP, Planck, LHC and ILC.}
  \label{fig:dmco}
\end{figure}

As already mentioned the lightest neutralino is a good candidate for the dark
matter particle. To calculate its density in the universe, the properties of
all particles contributing to the annihilation have to be reconstructed with
good precision. In any case the mixing angles and mass of the
$\tilde{\chi}^0_1$ need to be known. However also the properties of other
particles can be important. For example in the
$\tilde{\chi}^0_1-\tilde{\tau}_1$ coannihilation region the
$\tilde{\chi}^0_1-\tilde{\tau}_1$ mass difference is essential.  Figure
\ref{fig:dmco} shows the possible precision with which the dark matter density
and neutralino mass can be reconstructed from the LHC and the ILC
measurements\cite{dmfrancois}. 
ILC matches nicely the expected precision of the Planck satellite, allowing a
stringent test whether Supersymmetry can account for all dark matter in the
universe.

\section{Models without a Higgs}

Without a Higgs WW scattering becomes strong at high energy, finally violating
unitarity at 1.2\,TeV. One can thus expect new physics the latest at this
scale. At the moment there are mainly two classes of models that explain
electroweak symmetry breaking without a Higgs boson. In Technicolour like
models\cite{technicolour} new strong interactions are introduced at the TeV
scale. In Higgsless models the unitarity violation is postponed to higher
energy by new gauge bosons, typically KK excitations of the Standard Model
gauge bosons. Both classes should give visible signals at the ILC. The
accessible channels are W-pair production, where the exchanged $\gamma$ or Z
may fluctuate into a new state, vector boson scattering, where the new states
can be exchanged in the s- or t-channel of the scattering process and three
gauge boson production where the new states can appear in the decay of the
primary $\gamma$ or Z.

\subsection{Strong electroweak symmetry breaking}

\begin{figure}[htb]
  \centering
  \includegraphics[width=\linewidth,bb=0 0 546 513]{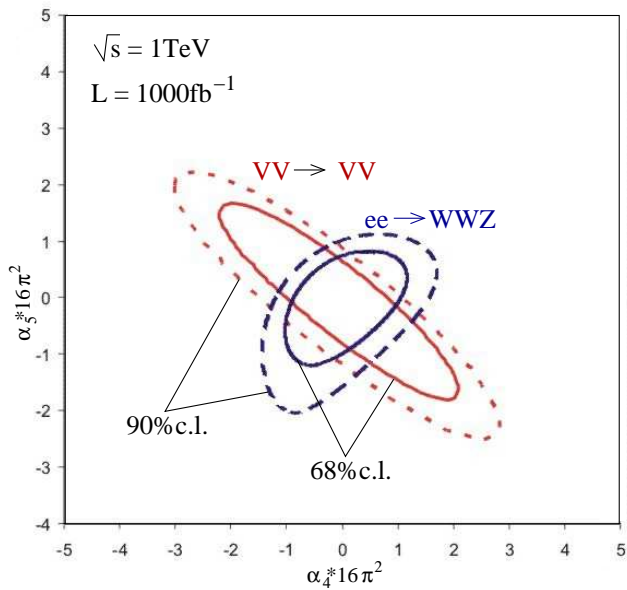}
  \caption{Expected sensitivity on $\alpha_4$ and $\alpha_5$ from vector boson
    scattering and three vector boson production.}
  \label{fig:predrag}
\end{figure}

As already said, in technicolour like models one expects new strong
interactions, including resonances, at the TeV scale. To analyse these models
in a model independent way, the triple and quartic couplings can be
parameterised by an effective Lagrangian in a dimensional
expansion\cite{wolfgang}. For the interpretation the effects of resonances on
these couplings can then be calculated. Figure \ref{fig:predrag} shows the
possible sensitivity to $\alpha_4$ and $\alpha_5$ at $\rts = 1\TeV$ from
vector-boson scattering and three vector boson production\cite{predrag}.
Typical sensitivities are ${\cal O}(0.1/16\pi^2)$ for triple and ${\cal
  O}(1/16\pi^2)$ for quartic couplings. This corresponds to mass limits around
$3\TeV$ for maximally coupled resonances. The different processes can then
distinguish between the different resonances. For example W-pair
production is only sensitive to vector resonances.

\subsection{Higgsless models}
\begin{figure}[htb]
  \centering
  \includegraphics[width=\linewidth]{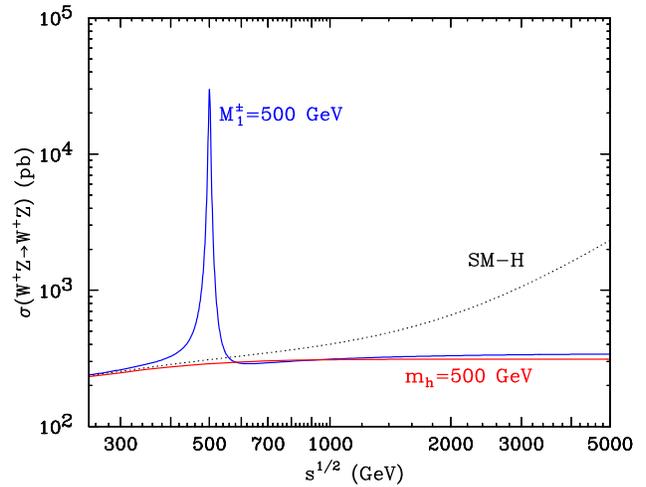}
  \caption[]{Cross section $\sigma(WZ \rightarrow WZ)$ in a Higgsless model
    and in the Standard Model with and without a Higgs\cite{higgsless2}.}
  \label{fig:hless}
\end{figure}

Higgsless models predict new gauge bosons at higher energies. Especially also
charged states are predicted that cannot be confused with a heavy Higgs.
Figure \ref{fig:hless} shows the cross section for the process $WZ \rightarrow
WZ$ in a Higgsless model, the Standard Model without a Higgs and the SM where
unitarity is restored by a 600\,GeV Higgs\cite{higgsless,higgsless2}.  Detailed
studies show that these states can be seen at LHC, however it is out of
question that such a state would also give a signal at ILC in $WZ \rightarrow
WZ$ and in $WWZ$ production so that its properties could be measured in
detail.

\section{Extra Gauge Bosons}
The ILC is sensitive to new gauge bosons in $\ee \rightarrow \ff$ via the
interference with the Standard Model amplitude far beyond $\sqrt{s}$.  The
sensitivity is typically even larger than at the LHC.  If the LHC measures the
mass of a new Z' a precise coupling measurement is possible at the ILC.  In
addition angular distributions are sensitive to the spin of the new state and
can thus distinguish for example between a Z' and KK graviton towers.
A review of the sensitivity can be found in\cite{phys_tdr}.

\begin{figure}[htb]
  \centering
  \includegraphics[width=0.9\linewidth]{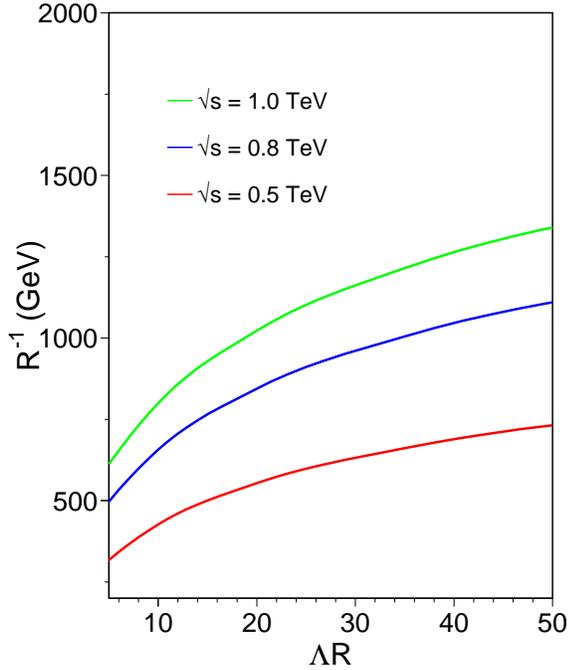}
  \caption[]{ILC reach for the $Z'',\,\gamma''$ expressed in $1/R$ as a 
    function of the cutoff parameter $\Lambda R$\cite{sabineed}. 
    A $\Lambda R$ value around 30 is suggested theoretically\cite{kkdmrc}.}
  \label{fig:kkdm}
\end{figure}

An interesting possibility is the reconstruction of the 2nd excitation of the
Z and $\gamma$ in universal extra dimensions.
In this models an excitation quantum number may be defined that is conserved
and makes the
lightest excitation stable and thus a good dark matter candidate\cite{kkdm}.
The second excitations couple to Standard Model particles only loop suppressed
and thus weakly\cite{kkdmrc}. Cosmology suggests 
$\frac{1}{R}\approx m(\gamma') < 1\TeV$ corresponding to 
$m(\gamma'') < 2\TeV$\cite{kkdm}. The LHC can see the $\gamma'$ in pair
production up to about this energy. The ILC is sensitive to the $Z''$ and
$\gamma''$ up to $2\rts$ which corresponds to the same $1/R$ reach for 
$\rts = 1\TeV$ (see fig.~\ref{fig:kkdm})\cite{sabineed}, helping enormously
in the interpretation of a possible LHC signal as KK excitation.

\begin{figure}[htb]
  \centering
  \includegraphics[width=\linewidth]{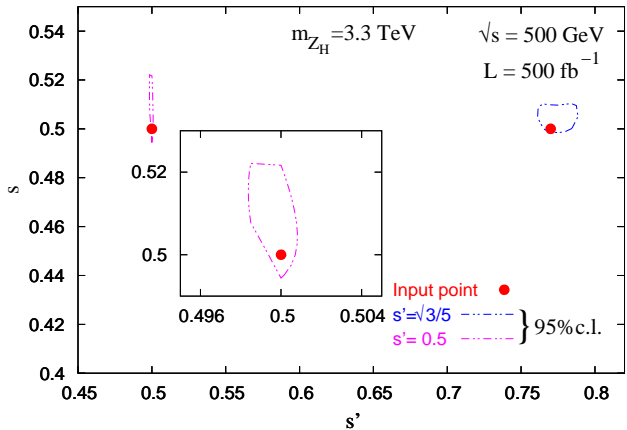}
  \caption{Measurement of the $Z_H$ mixing angles at the ILC.}
  \label{fig:littleh}
\end{figure}

Little Higgs models explain the ``little hierarchy problem'' by a new gauge
structure and a new top-like quark\cite{littleh}.  The new gauge structure
also predicts new vector bosons ($Z_H,\, A_H,\, W_H$) at masses of a few TeV.
Figure \ref{fig:littleh} shows the precision with which the mixing angles of
the $Z_H$ can be measured at $\rts = 500 \GeV$ once its mass ($3.3\TeV$ in
this example) in measured at the LHC\cite{conleylh}.

\section{Conclusions}
Independent of which physics scenario nature has chosen, the ILC will be
needed in addition to the LHC. If there is a Higgs and SUSY the ILC has to
reconstruct as many of the SUSY-breaking parameters as possible, extrapolate
them to the GUT scale to get some understanding of the breaking mechanism and
measure the properties of the dark matter particle.

If there is a Higgs without Supersymmetry the precision measurements of the
Higgs boson guide the way to the model of electroweak symmetry breaking. In
addition several models, like some extra dimension models or little Higgs
models have extra gauge bosons that are visible via their indirect effects.

If the LHC doesn't find any Higgs boson, the ILC can fill some loopholes that
still exist, can see signals of strong electroweak symmetry breaking and is
sensitive to a new gauge sector.

In any case we know that the top quark is accessible to the ILC and that its
properties can be measured with great precision.

 \section*{Acknowledgements}
 I would like to thank everybody who helped me in the preparation of this talk,
 especially Klaus Desch, Jonathan Feng, Fabiola Gianotti, Francois Richard,
 Sabine Riemann and Satoru Yamashita. Special thanks also to Tord Ekelof and
 the organising committee for the splendid organisation and hospitality during
 the conference!

%%%%%%%%%%%%%%%%%%%%%%%%%%%%%%%%
%  Question and Answer Section %
%%%%%%%%%%%%%%%%%%%%%%%%%%%%%%%%
% Use clear page to make sure everything is flush and a new
% page is started (not just a new column)
%%%%%%%%%%%%%%%%%%%%%%%%%%%%%%%%
%\clearpage
\newpage
\section*{DISCUSSION}

\begin{description}
\item[Bernd Jantzen] (Univ. of Karlsruhe): Can anything be said about if we
  need CLIC and what we would like to explore with it before the data of LHC
  and/or ILC has been analysed?
  
\item[Klaus M\"onig{\rm :}] The detailed physics case for CLIC can only be
  made once we know the scenario realised in nature. For example if relatively
  light SUSY exists CLIC can extend the ILC precision measurements to the
  coloured part of the spectrum. However, it may also be possible, that a
  hadron collider at very high energy may be the better next machine at the
  energy frontier.

\end{description}

\end{document}

\end{document}

%% file: definitions.tex
\def\ifmath#1{\relax\ifmmode #1\else $#1$\fi}%
\def\TeV{\ifmmode {\,\mathrm{ Te\kern -0.1em V}}\else
                   \textrm{Te\kern -0.1em V}\fi}%
\def\GeV{\ifmmode {\,\mathrm{ Ge\kern -0.1em V}}\else
                   \textrm{Ge\kern -0.1em V}\fi}%
\def\MeV{\ifmmode {\,\mathrm{ Me\kern -0.1em V}}\else
                   \textrm{Me\kern -0.1em V}\fi}%
\def\keV{\ifmmode {\,\mathrm{ ke\kern -0.1em V}}\else
                   \textrm{ke\kern -0.1em V}\fi}%
\def\eV{\ifmmode  {\,\mathrm{ e\kern -0.1em V}}\else
                   \textrm{e\kern -0.1em V}\fi}%

\newcommand{\fb}{\,{\rm fb}}

\newcommand{\fbi}{\, \fb^{-1}}

\newcommand{\lunit}{\mathrm{cm}^{-2}\mathrm{s}^{-1}}

\newcommand{\ff}    {{\rm f}\overline{\rm f}}

\newcommand{\ee}    {\mathrm{e}^+\mathrm{e}^-}
\newcommand{\mumu}  {\mu^+\mu^-}
\newcommand{\ttb}   {{\mathrm t\bar{\mathrm t}}}
\newcommand{\bb}    {{\mathrm b\bar{\mathrm b}}}

\newcommand{\MH}      {m_{\mathrm{H}}}

\newcommand{\MA}      {m_{\mathrm{A}}}

\newcommand{\MT}      {m_{\mathrm{t}}}

\newcommand{\rts}   {\sqrt{s}}

%% file: title.tex
\begin{titlepage}

\pagenumbering{arabic}
\begin{flushright}
DESY 05-207\\
LAL 05-98\\
hep-ph/0509159\\
September 15, 2005
\end{flushright}
%========================================================================%
\vspace*{2.cm}
\begin{center}
\Large 
\boldmath
{\bf
%===================> DELPHI note title        =====> To be filled <=====%
Physics at Future Linear Colliders
%========================================================================%
} \\
\unboldmath
\vspace*{2.cm}
\normalsize { 
%===================> DELPHI note author list  =====> To be filled <=====%
   {\bf K. M\"onig}\\
   {\footnotesize DESY, Zeuthen, Germany and LAL, Orsay France
     \\E-mail: klaus.moenig@desy.de
   }\\
   
%========================================================================%
}
%\vspace*{2.cm}
\end{center}
\vspace{\fill}
\begin{abstract}
\noindent
This article summarises the physics at future linear colliders. It will be
shown that in all studied physics scenarios a 1\,TeV linear collider in
addition to the LHC will enhance our knowledge significantly and helps to 
reconstruct the model of new physics nature has chosen.
\end{abstract}
\vspace{\fill}
\begin{center}
%==========> Proceedings.. presented at ..==> To be filled if needed<=====%
Invited talk at the Lepton Photon Symposium 2005, Upsala, Sweden, July 2005
%=========================================================================%
\end{center}
\vspace{\fill}
\end{titlepage}

%%% Local Variables: 
%%% mode: latex
%%% TeX-master: "lp_lc"
%%% End: 